\documentclass{Interspeech2024}
\usepackage{scrextend}

\interspeechcameraready

\title{Multi-modal Adversarial Training for Zero-Shot Voice Cloning}

\name{John Janiczek, Dading Chong, Dongyang Dai, Arlo Faria$^1$, \\Chao Wang, Tao Wang, Yuzong Liu}

\address{
  Zoom Video Communications, United States of America}
\email{john.janiczek@zoom.us, yuzong.liu@zoom.us}


\begin{document}

\maketitle

\begin{abstract}
    
A text-to-speech (TTS) model trained to reconstruct speech given text tends towards predictions that are close to the average characteristics of a dataset, failing to model the variations that make human speech sound natural.
 This problem is magnified for zero-shot voice cloning, a task that requires training data with high variance in speaking styles. 
 We build off of recent works which have used Generative Advsarial Networks (GAN) by proposing a Transformer encoder-decoder architecture to conditionally discriminates between real and generated speech features. 
 The discriminator is used in a training pipeline that improves both the acoustic and prosodic features of a TTS model. 
 We introduce our novel adversarial training technique by applying it to a FastSpeech2 acoustic model and training on Libriheavy, a large multi-speaker dataset, for the task of zero-shot voice cloning. 
 Our model achieves improvements over the baseline in terms of speech quality and speaker similarity. Audio examples from our system are available online$^2$.
\end{abstract}

\let\thefootnote\relax\footnote{$^1$Now affiliated with Cartesia AI.}
\let\thefootnote\relax\footnote{$^2$ \url{https://johnjaniczek.github.io/m2gan-tts/}}

\section{Introduction}

The goal of text-to-speech (TTS) is to generate human-like speech from a given text input. Neural networks have been shown to advance the state of the art in TTS by synthesizing speech that nearly matches the naturalness of human speech \cite{tan2024naturalspeech}. Recently, there has been an interest in extending neural TTS with advanced features such as zero-shot voice cloning. The purpose of this feature is to allow the TTS to synthesize speech in the voice of a target speaker. These systems implement voice cloning by encoding speech from a short reference audio segment, and they do not require any additional training or fine-tuning. \cite{jia2018transfer} \cite{zhang2020gazev} 
\cite{casanova2022yourtts} \cite{wang2023neural} \cite{zhang2023speak} \cite{yang2021ganspeech} \cite{lee2021multi}. It is a challenging problem, because the neural network must extrapolate to unseen speakers - and there is still a quality gap between zero-shot and few-shot systems.

One of the core problems in achieving natural and expressive neural TTS is the one-to-many mapping between text and speech. TTS models trained with a reconstruction loss tend to predict the expected value which can cause TTS systems to make over-smooth predictions that are close to the average behavior of the dataset rather than capturing the diversity of human speech \cite{ren2022revisiting}.  We hypothesize that part of the reason that there is a quality gap in zero-shot voice cloning, is that the large multi-speaker datasets used to train these models adds even more variance in this one-to-many mapping because of all the different voices and speaking styles in the dataset.

We first look at previous approaches which have been used to improve over-smooth TTS predictions. In \cite{ren2020fastspeech} it was shown that this one to many problem can be mitigated by conditioning the acoustic decoder on ground truth prosodic features which removes the amount of unknown variables in the one-to-many mapping. Building off of this, other works propose letting the model learn to encode the relevant prosodic features \cite{liu2021delightfultts} \cite{tan2024naturalspeech} \cite{kim2021conditional}. While these approaches alleviate the one-to-many mapping in the decoder, these hidden prosodic features must be predicted by the model which is yet another one-to-many mapping and during inference the resulting speech can still sound inexpressive in comparison to human speech especially when extended to a large number of speakers.

Other recent works have tried to change the training paradigm to improve the naturalness and expressiveness of TTS. For example, Diffusion based TTS models learn to de-noise speech \cite{popov2021grad} \cite{gao2023e3} or de-noise prosodic features \cite{li2023diverse}. Since these models learn to denoise in small steps, they do not have the same problems of systems which are trained to reconstruct speech in a single pass. However, it usually takes many denoising steps to produce high quality audio and these systems are too slow for some applications \cite{tan2021survey}. Motivated by producing high quality audio without the resource constraints of Diffusion we explore other architectures.

In this paper we focus on using a generative adversarial network (GAN) to mitigate the one-to-many mapping problem. We consider our TTS acoustic model as the generator, which produces both acoustic features and prosodic features. A separate discriminator model is trained to distinguish between the synthesized features and the ground truth features. As the discriminator learns to differentiate real data from synthetic data, it can then be used to guide the optimization of the synthetic data with an adversarial loss so that the generated features better match the distribution of real data \cite{binkowski2019high} \cite{kim2021conditional} \cite{kong2023vits2} \cite{zhang2020gazev}. This has previously been shown to reduce the over-smoothing problem in TTS systems. \cite{ren2022revisiting} \cite{sheng2019reducing}.

In this paper we experiment with using a more expressive discriminator than previous techniques so that we can scale up the GAN-based training to address the quality gap that exists in zero-shot voice cloning TTS. In specific we chose a Transformer because it has been shown to have high performance when scaling up data and parameters. Our main contributions are as follows:

\begin{itemize}
    \item A novel design for a Transformer encoder-decoder discriminator which uses Multi-modal Fusion to conditionally discriminate TTS features.
    \item A Multi-feature Generative Adversarial Training pipeline which uses our discriminator to enhance both acoustic and prosodic features for natural and expressive TTS.
    \item Extensive experiments with a large multi-speaker dataset (50,000 hours) to produce high quality and expressive zero-shot voice cloning with a light-weight model that can run in real-time on a CPU.
\end{itemize}

\section{Methodology}


To effectively address the issues of prosody smoothing and weak transferability caused by one-to-many mapping in TTS, this section will first provide a brief overview of the training framework. Subsequently, it will focus on introducing the innovations of this paper, including the Multi-modal Fusion Discriminator network and the Multi-feature Generative Adversarial Training approach.

\subsection{Training Framework}

\begin{figure}[t]
  \centering
  \includegraphics[width=\linewidth]{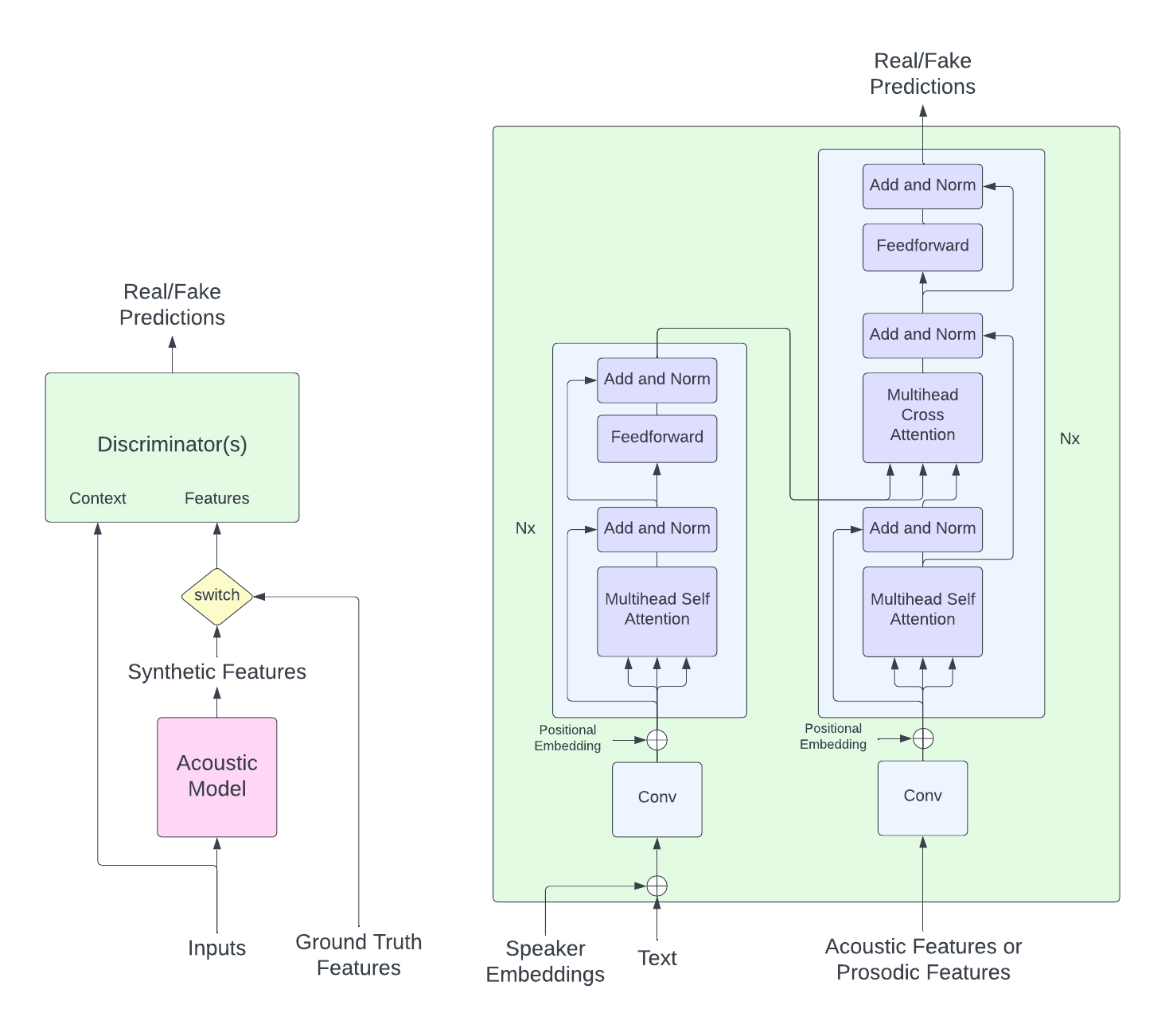}
  \caption{The TTS acoustic model is trained with a generative adversarial network (left). The discriminator is a Transformer (right) that uses the text and speaker embedding as conditional inputs to determine whether the TTS features are real or fake. In our experiments we train separate Transformers which discriminate against acoustic features or prosodic features.}
  \label{fig:system_architecture}
\end{figure}

The overall training framework of the model is illustrated on the left side of Figure \ref{fig:system_architecture}, which follows a Generative Adversarial modeling approach. To address the issue of over-smooth synthesized acoustic features caused by the one-to-many mapping problem in the TTS task, we employ a discriminator to distinguish between the acoustic features predicted by the acoustic model and the ground truth acoustic features and an adversarial loss to guide the optimization of the acoustic feature parameters. It is worth mentioning that the utilization of a discriminator for adversarial training of the acoustic model has been proposed in some previous works. \cite{sheng2019reducing} \cite{ren2022revisiting} \cite{kim2021conditional} \cite{kong2020hifi} \cite{binkowski2019high} \cite{zhang2020gazev} \cite{kong2023vits2} However, unlike prior approaches, we conduct in-depth analysis and experiments on the discriminator structure, proposing a Multi-modal Fusion Discriminator Network and a Multi-feature Generative Adversarial Training approach to enhance the discriminator's capability. This, in turn, improves the quality and transferability of the synthesized speech generated by the acoustic model when applied to the problem of zero-shot voice cloning.

\subsection{Multi-modal Fusion Discriminator Network}
The purpose of the discriminator is to encourage the acoustic model to generate more diverse information and avoid over-smoothing issues. However, solely judging the authenticity of acoustic features at the feature level may lead the discriminator to prioritize making the feature distribution as close as possible to that of real speech, overlooking any differentiation in lexical content, speaker identity, and other contextual information. This is contradictory to the original intention of designing the discriminator to make the acoustic features as diverse as possible.
Therefore, when determining the authenticity of acoustic features, it is essential to consider not only the feature level but also incorporate additional modalities of contextual information to aid discrimination.

For instance, lexical information can be leveraged, as the acoustic features of a frame depend on the semantic content of that frame's speech. If the lexical information is excluded, judging solely from the acoustic feature level may result in identical acoustic features for different semantics, leading to a loss of differentiation. Additionally, speaker information serves as crucial prior knowledge for acoustic features. Without considering speaker information, the discriminator may only judge whether the acoustic features are smooth, but by incorporating speaker information, the discriminator can also learn the correlation between acoustic features and speakers, further enhancing the diversity of synthesized acoustic features.

Based on this hypothesis, this work aims to integrate various contextual information into the discriminator. Due to the varying dimensions of different modalities, we propose a Transformer-based encoder-decoder discriminator model, as illustrated on the right side of Figure \ref{fig:system_architecture}. The encoder part takes input text information and speaker information, while the decoder part takes input feature information to be discriminated and predicts the authenticity of the feature information. By adopting this approach, it can assist the discriminator in evaluating acoustic features more comprehensively, thereby enhancing the diversity of acoustic features.

\subsection{Multi-feature Generative Adversarial Training}
In the previous section, we elaborated on enhancing the diversity of features predicted by the acoustic model by designing a discriminator with additional contextual modalities. To further enhance the diversity of acoustic features, this paper conducts adversarial training on both acoustic and prosodic features. The overall training process is as follows. 

Assuming acoustic features (such as mel spectrograms) predicted by the acoustic model are denoted as $\hat{y}_a$ and ground truth acoustic features are denoted as $y_a$, the generative loss is 
\begin{equation}
\mathcal{L}_{G_a} = \text{MAE}(\hat{y}_a, y_a)
\end{equation}
where $\text{MAE}$ is the mean absolute error. 

The discriminator $D_a$ is also used to guide the optimization of the acoustic model using an adversarial loss:
\begin{equation}
\mathcal{L}_{A_a} = -D_a(\hat{y}_a|x_t, x_s)    
\end{equation}
 where $x_t$ represents the text information input to the discriminator, while $x_s$ denotes the speaker representation.
 
The optimization loss for the acoustic feature discriminator is based on a conditional hinge loss \cite{kavalerov2021multi}. Note that in practice our discriminator produces a prediction for each frame, and the hinge loss is applied to each output:
\begin{equation}
\begin{aligned}
\mathcal{L}_{D_a} = &-\min(0,D_a(y_a|x_t, x_s) - 1) \\
&- \min(0, -D_a(\hat{y}_a|x_t, x_s) - 1)
\end{aligned}
\end{equation}

In addition to supervising the acoustic features, we also conduct adversarial training on predicted prosodic features to address the potential issue of overly smooth prosodic features during inference. 
These prosodic features may comprise pitch, energy, duration, and encoded prosodic embeddings. 
Let $\hat{y}_p$ denote the prosodic features generated by the model, and $y_p$ denote the corresponding ground truth prosodic features.
(Ground truth audio is used for encoding the ground truth prosodic embeddings, using the encoder from \cite{liu2021delightfultts}.)
Accordingly, the acoustic model's generative optimization loss is
\begin{equation}
\mathcal{L}_{G_p} = \text{MSE}(\hat{y}_p, y_p)
\end{equation}
where $\text{MSE}$ is the mean squared error. 
The optimization losses involving the prosodic feature discriminator are
\begin{equation}
\mathcal{L}_{A_p} = -D_p(\hat{y}_p|x_t, x_s)
\end{equation}
\begin{equation}
\begin{aligned}
\mathcal{L}_{D_p} = &-\min(0,D_p(y_p|x_t, x_s) - 1) \\
&- \min(0, -D_p(\hat{y}_p|x_t, x_s) - 1) \\
\end{aligned}
\end{equation}
where $D_p$ represents the discriminator for the prosodic features. 

In summary, the optimization loss for the acoustic model is 
\begin{equation}
\label{eq:total_ga}
\mathcal{L}_{GA} = \mathcal{L}_{G_a}+ \mathcal{L}_{G_p} + \lambda_{A}(\mathcal{L}_{A_a} + \mathcal{L}_{A_p})    
\end{equation}
while the total discriminator loss is given by:
\begin{equation}
\label{eq:total_d}
\mathcal{L}_{D} = \mathcal{L}_{D_a}  + \mathcal{L}_{D_p} 
\end{equation}
The $\lambda_{A}$ term is introduced as a hyperparameter to scale the adversarial loss.
Importantly, only parameters of the acoustic model are updated for $\mathcal{L}_{GA}$ and only parameters of the discriminators are updated for $\mathcal{L}_{D}$ since these models have competing objectives.

\section{Experiments}

\subsection{Dataset details}

Our experiments use the Libriheavy \cite{kang2023libriheavy} and LibriTTS-R \cite{koizumi2023libritts} datasets. The Libriheavy large training subset contains over 50,000 hours of speech from nearly 7,000 speakers, providing a diverse range of speaking styles. We notice that Libriheavy has transcription errors, especially with numeric text (e.g. unspoken page numbers and footnotes), so we filter out all utterances containing  numerals. This removes almost 7 percent of the data, but resolves an issue where the model would otherwise skip over digit sequences during inference. Since we ideally want our model to be capable of producing studio-quality speech, we also include the LibriTTS-R dataset which contains ~500 hours of speech which has been restored to near studio quality. 

Our evaluations use the LibriTTS-R test dataset for evaluating our performance on unseen speakers. The holdout datasets for Libriheavy and LibriTTS-R have been split so that the speakers in the test dataset have never been seen during training. For each testing example, there is an associated ground truth of the target speaker saying the utterance. However, the reference audio that is passed to the speaker encoder is always sampled from a different utterance from the same speaker to better simulate a realistic zero-shot voice cloning scenario.

\subsection{Model and training details}
Our baseline model is based off of FastSpeech2 \cite{ren2020fastspeech} with the following modifications. Firstly, we add the prosodic encoder from \cite{liu2021delightfultts} which we find is needed as pitch, energy, and duration do not adequately capture the differences in speaking style over our large training dataset. Secondly, we add a speaker encoder which has been pretrained on a speaker identification task and is frozen during training to enable zero-shot voice cloning as described in \cite{jia2018transfer}. The acoustic model contains 45M parameters. We also use a HifiGAN \cite{kong2020hifi} vocoder to synthesize the FastSpeech2 spectrograms into 24kHz audio.

Our proposed discriminator model is an encoder-decoder Transformer with additional convolutional layers at the input. The hyperparameters for our discriminator are shown in Table \ref{tab:hyperparams}. Note that the acoustic discriminator has a stride of 2 in the convolutional layers to downsample the features; this can reduce the memory constraint of our training by reducing the sequence length of our audio frames prior to passing it to the acoustic decoder. We find that the downsampling does not degrade the performance of the descriminator for the proposed configuration. In addition, note that the prosody discriminator needs to fuse several prosodic features together. Each prosodic feature has a separate convolutional layer which projects the features into the same dimensionality before summing the features together with the positional encoding at the transformer input. We add a fixed bias of 10.0 along the diagonal of cross-attention energies; this guides alignment of our discriminators and speeds convergence.

\begin{table}[th]
  \caption{Discriminator Hyperparameters}
  \label{tab:hyperparams}
  \centering
  \begin{tabular}{ll}
    \toprule
    \textbf{Hyperparameter} & \textbf{Value} \\
    \midrule
        $\lambda_{A}$ & 0.1 \\
        Convolution layers & 2 \\
        Convolution kernel size & 11 \\
        Convolution stride & 2 (acoustic) \\
            & 1 (prosodic) \\
        Transformer encoder layers & 2 \\
        Transformer decoder layers & 6 \\
        Transformer hidden dim & 512 (acoustic) \\
         & 256 (prosodic) \\
        Transformer feedforward dim & 1024 (acoustic)\\
         & 512 (prosodic) \\
        Transformer heads & 4 \\
        Transformer dropout & 0.1 \\
    \bottomrule
  \end{tabular}
  
\end{table}

For all models, we train for 3 epochs using 8 H100 GPUs. We use a batch sampler which batches audio samples with similar lengths together to reduce padding and samples batches with a maximum of 8,000 audio frames per GPU. 

We use the AdamW \cite{loshchilov2017decoupled} optimizer with a learning rate of $0.002$, a weight decay of $0.01$, $\beta_1 = 0.5$ and $\beta_2 = 0.9$. For the first epoch, the discriminators are trained to distinguish between synthetic and real data but the adversarial loss is not applied. The adversarial losses for the acoustic and prosodic discriminators are added in the second and third epoch, respectively. We use the learning rate schedule as described in \cite{vaswani2017attention} and we restart the learning rate warm-up at the beginning of each epoch. This training schedule was based on the observation that when an adversarial loss is introduced to the generator it can potentially cause the model to change too rapidly, so we take care to slowly introduce the influence of the adversarial loss.

\subsection{Metrics}

We use three objective metrics to evaluate the model's audio quality, speaker similarity, and expressiveness. The NISQA MOS \cite{mittag2021nisqa} metric uses a model which has been trained to predict human preference and has been shown to correlate with quality of speech when averaged over a sufficient number of samples. We also use a speaker similarity metric which takes the cosine similarity between speaker embeddings from the synthesized speech and the reference audio. For the speaker similarity metric we use an off-the-shelf speaker encoder \cite{ravanelli2021speechbrain} to extract speaker embeddings. We use a different separate speaker encoder for evaluation because we have noticed that speaker encoders tend to give higher scores when the acoustic model is trained on the same speaker encoder. As proxy for determining the expressiveness of speech, we measure the pitch standard deviation similar to \cite{ren2020fastspeech} to see how the variation in pitch compares between ground truth and synthesized examples. The pitch standard deviation is calculated for each utterance and then averaged across the test set, so that the variation is dominated by differences in pitch within the utterance rather than differences in pitch among various different speakers.

\subsection{Model evaluation and ablation studies}
We first evaluate our GAN-based approach by comparing it directly against the baseline model and keeping all training parameters the same other than the introduction of the discriminator. We also show the metrics for both the reference and ground truth since the reference audio which is passed to the speaker encoder of the model is always sampled from a different utterance than the ground truth example. As shown in Table \ref{tab:evaluation}
 the proposed model shows significant performance improvements over the baseline model with higher NISQA MOS and speaker similarity scores. The pitch standard deviation is also closer to the reference audio indicating that the prosody of the proposed model is more expressive. 

\begin{table}[th]
  \caption{Evaluation of the proposed model on the task of synthesizing unseen speakers from the LibriTTS-R dataset.}
  \label{tab:evaluation}
  \centering
  \scalebox{0.85}{
  \begin{tabular}{lccc}
    \toprule
     & \textbf{NISQA MOS  $\uparrow$} & \textbf{Speaker Sim. $\uparrow$} & \textbf{Pitch Std.} \\
    \midrule
        Reference &  4.30 & 1.00 & 34.00 \\
        Ground Truth &  4.26 & 0.57 & 35.19 \\
    \midrule
        Baseline  &	3.27 & 0.48 & 25.10 \\
        Proposed  & \textbf{4.15} & \textbf{0.55} & \textbf{33.55} \\
    \bottomrule
  \end{tabular}}
  
\end{table}

Next we show several ablation studies to demonstrate the significance of each component of our discriminator. The results of the ablation studies are shown in Table 
\ref{tab:ablation}.  

In the first ablation study, conditional inputs are removed from the discriminator to show the effect of passing the text and speaker embedding to the discriminator. Since this ablation study removes the Transformer conditional encoder entirely we increase the Transformer layers to 8 to match the parameter count of the encoder-decoder model for a more fair comparison. This experiment has worse quality across all metrics showing the importance of context for our Multi-modal Fusion Discriminator network.

In the next ablation study we remove the speaker embedding from the Transformer discriminator and only condition on text. This experiment has worse overall quality showing the importance of the speaker embedding condition in our design.

We also run an ablation study where we only use the acoustic discriminator without the prosodic discriminator. This model does not affect the speaker similarity, but the pitch standard deviation is much further from the reference audio, showing the importance of the prosodic discriminator at encouraging expressive speech for our Multi-feature Generative Adversarial Training approach.

Finally we compare different Transformer architectures to explain the decisions behind our proposed design, as shown in Table \ref{tab:architectures}. For the encoder-only architecture we simply pass a concatenated sequence of features to the Transformer encoder. The decoder is bypassed in this configuration, but the encoder is increased to 8 layers to keep the parameter count similar for this comparison. We find that while the encoder-only architecture can also be used to improve over the baseline, the encoder-decoder architecture gives us better performance. We do not consider a decoder-only architecture as we find that causal masking does not help for our training pipeline. We also compare a balanced architecture with 4 encoder layers and 4 decoder layers against a decoder-heavy architecture with 2 encoder layers and 6 decoder layers. We find that while the balanced architecture has a slightly better speaker similarity, the decoder-heavy model has significantly better NISQA MOS and pitch standard  deviation. Because the decoder-heavy architecture performs the best overall we use this as our proposed design in all other experiments.

\begin{table}[th]
  \caption{Ablation tests showing the effect of removing conditional inputs and the prosodic discriminator.}
  \label{tab:ablation}
  \centering
  \scalebox{0.85}{
  \begin{tabular}{lccc}
    \toprule
    \textbf{Model} & \textbf{NISQA MOS  $\uparrow$} & \textbf{Speaker Sim. $\uparrow$} & \textbf{Pitch Std.} \\
    \midrule
        w/o text \& spkr & -0.12 & -0.01 & -2.65 \\
        w/o speaker  &	-0.27 & -0.01 & -5.56 \\
        w/o prosody  & -0.13 & 	-0.00 & 	-8.87 \\
    \bottomrule
  \end{tabular}}
  
\end{table}

\begin{table}[th]
  \caption{Ablation tests comparing Transformer architectures.}
  \label{tab:architectures}
  \centering
    \scalebox{0.85}{
  \begin{tabular}{lccc}
    \toprule
    \textbf{Model} & \textbf{NISQA MOS  $\uparrow$} & \textbf{Speaker Sim. $\uparrow$} & \textbf{Pitch Std.} \\
    \midrule
        Enc-Dec. 2:6  & \textbf{4.15} & 0.55 & \textbf{33.55} \\
        Enc-Dec 4:4   & 4.05 & \textbf{0.56} & 30.23 \\
        Enc-Only 8:0   & 3.94 &  0.55 & 25.98\\
    \bottomrule
  \end{tabular}}
  
\end{table}

\section{Conclusion and Discussion}

In this work, we propose a Multi-modal Fusion Discriminator and Multi-feature Generative Adversarial Training. The proposed technique is capable of encouraging the acoustic model to produce diverse, expressive, and natural-sounding speech because of its ability to discriminate if the model matches the lexical and speaker-dependent characteristics of the synthesized speech. We show that our proposed design makes significant improvements over a baseline model on the task of zero-shot voice cloning. We also run extensive ablation studies to demonstrate the significance of each of our the design decisions.

One disadvantage of the GAN-based approach is that it adds complexity to the training pipeline and care must be taken to ensure that the proposed approach converges. Nevertheless, during inference there is no additional complexity or latency and the proposed FastSpeech2 model can be deployed on a CPU with real-time performance \cite{ren2020fastspeech}. We hope that this work enables high-quality zero-shot voice cloning for applications where latency and resources are constrained.

\bibliographystyle{IEEEtran}
\bibliography{mybib}

\begin{thebibliography}{10}
\providecommand{\url}[1]{#1}
\csname url@samestyle\endcsname
\providecommand{\newblock}{\relax}
\providecommand{\bibinfo}[2]{#2}
\providecommand{\BIBentrySTDinterwordspacing}{\spaceskip=0pt\relax}
\providecommand{\BIBentryALTinterwordstretchfactor}{4}
\providecommand{\BIBentryALTinterwordspacing}{\spaceskip=\fontdimen2\font plus
\BIBentryALTinterwordstretchfactor\fontdimen3\font minus \fontdimen4\font\relax}
\providecommand{\BIBforeignlanguage}[2]{{%
\expandafter\ifx\csname l@#1\endcsname\relax
\typeout{** WARNING: IEEEtran.bst: No hyphenation pattern has been}%
\typeout{** loaded for the language `#1'. Using the pattern for}%
\typeout{** the default language instead.}%
\else
\language=\csname l@#1\endcsname
\fi
#2}}
\providecommand{\BIBdecl}{\relax}
\BIBdecl

\bibitem{tan2024naturalspeech}
X.~Tan, J.~Chen, H.~Liu, J.~Cong, C.~Zhang, Y.~Liu, X.~Wang, Y.~Leng, Y.~Yi, L.~He \emph{et~al.}, ``Naturalspeech: End-to-end text-to-speech synthesis with human-level quality,'' \emph{IEEE Transactions on Pattern Analysis and Machine Intelligence}, 2024.

\bibitem{jia2018transfer}
Y.~Jia, Y.~Zhang, R.~Weiss, Q.~Wang, J.~Shen, F.~Ren, P.~Nguyen, R.~Pang, I.~Lopez~Moreno, Y.~Wu \emph{et~al.}, ``Transfer learning from speaker verification to multispeaker text-to-speech synthesis,'' \emph{Advances in neural information processing systems}, vol.~31, 2018.

\bibitem{zhang2020gazev}
Z.~Zhang, B.~He, and Z.~Zhang, ``Gazev: Gan-based zero-shot voice conversion over non-parallel speech corpus,'' \emph{arXiv preprint arXiv:2010.12788}, 2020.

\bibitem{casanova2022yourtts}
E.~Casanova, J.~Weber, C.~D. Shulby, A.~C. Junior, E.~G{\"o}lge, and M.~A. Ponti, ``Yourtts: Towards zero-shot multi-speaker tts and zero-shot voice conversion for everyone,'' in \emph{International Conference on Machine Learning}.\hskip 1em plus 0.5em minus 0.4em\relax PMLR, 2022, pp. 2709--2720.

\bibitem{wang2023neural}
C.~Wang, S.~Chen, Y.~Wu, Z.~Zhang, L.~Zhou, S.~Liu, Z.~Chen, Y.~Liu, H.~Wang, J.~Li \emph{et~al.}, ``Neural codec language models are zero-shot text to speech synthesizers,'' \emph{arXiv preprint arXiv:2301.02111}, 2023.

\bibitem{zhang2023speak}
Z.~Zhang, L.~Zhou, C.~Wang, S.~Chen, Y.~Wu, S.~Liu, Z.~Chen, Y.~Liu, H.~Wang, J.~Li \emph{et~al.}, ``Speak foreign languages with your own voice: Cross-lingual neural codec language modeling,'' \emph{arXiv preprint arXiv:2303.03926}, 2023.

\bibitem{yang2021ganspeech}
J.~Yang, J.-S. Bae, T.~Bak, Y.~Kim, and H.-Y. Cho, ``Ganspeech: Adversarial training for high-fidelity multi-speaker speech synthesis,'' \emph{arXiv preprint arXiv:2106.15153}, 2021.

\bibitem{lee2021multi}
S.-H. Lee, H.-W. Yoon, H.-R. Noh, J.-H. Kim, and S.-W. Lee, ``Multi-spectrogan: High-diversity and high-fidelity spectrogram generation with adversarial style combination for speech synthesis,'' in \emph{Proceedings of the AAAI Conference on Artificial Intelligence}, vol.~35, no.~14, 2021, pp. 13\,198--13\,206.

\bibitem{ren2022revisiting}
Y.~Ren, X.~Tan, T.~Qin, Z.~Zhao, and T.-Y. Liu, ``Revisiting over-smoothness in text to speech,'' \emph{arXiv preprint arXiv:2202.13066}, 2022.

\bibitem{ren2020fastspeech}
Y.~Ren, C.~Hu, X.~Tan, T.~Qin, S.~Zhao, Z.~Zhao, and T.-Y. Liu, ``Fastspeech 2: Fast and high-quality end-to-end text to speech,'' \emph{arXiv preprint arXiv:2006.04558}, 2020.

\bibitem{liu2021delightfultts}
Y.~Liu, Z.~Xu, G.~Wang, K.~Chen, B.~Li, X.~Tan, J.~Li, L.~He, and S.~Zhao, ``Delightfultts: The microsoft speech synthesis system for blizzard challenge 2021,'' \emph{arXiv preprint arXiv:2110.12612}, 2021.

\bibitem{kim2021conditional}
J.~Kim, J.~Kong, and J.~Son, ``Conditional variational autoencoder with adversarial learning for end-to-end text-to-speech,'' in \emph{International Conference on Machine Learning}.\hskip 1em plus 0.5em minus 0.4em\relax PMLR, 2021, pp. 5530--5540.

\bibitem{popov2021grad}
V.~Popov, I.~Vovk, V.~Gogoryan, T.~Sadekova, and M.~Kudinov, ``Grad-tts: A diffusion probabilistic model for text-to-speech,'' in \emph{International Conference on Machine Learning}.\hskip 1em plus 0.5em minus 0.4em\relax PMLR, 2021, pp. 8599--8608.

\bibitem{gao2023e3}
Y.~Gao, N.~Morioka, Y.~Zhang, and N.~Chen, ``E3 tts: Easy end-to-end diffusion-based text to speech,'' in \emph{2023 IEEE Automatic Speech Recognition and Understanding Workshop (ASRU)}.\hskip 1em plus 0.5em minus 0.4em\relax IEEE, 2023, pp. 1--8.

\bibitem{li2023diverse}
X.~Li, S.~Liu, M.~W. Lam, Z.~Wu, C.~Weng, and H.~Meng, ``Diverse and expressive speech prosody prediction with denoising diffusion probabilistic model,'' \emph{arXiv preprint arXiv:2305.16749}, 2023.

\bibitem{tan2021survey}
X.~Tan, T.~Qin, F.~Soong, and T.-Y. Liu, ``A survey on neural speech synthesis,'' \emph{arXiv preprint arXiv:2106.15561}, 2021.

\bibitem{binkowski2019high}
M.~Bi{\'n}kowski, J.~Donahue, S.~Dieleman, A.~Clark, E.~Elsen, N.~Casagrande, L.~C. Cobo, and K.~Simonyan, ``High fidelity speech synthesis with adversarial networks,'' \emph{arXiv preprint arXiv:1909.11646}, 2019.

\bibitem{kong2023vits2}
J.~Kong, J.~Park, B.~Kim, J.~Kim, D.~Kong, and S.~Kim, ``Vits2: Improving quality and efficiency of single-stage text-to-speech with adversarial learning and architecture design,'' \emph{arXiv preprint arXiv:2307.16430}, 2023.

\bibitem{sheng2019reducing}
L.~Sheng and E.~N. Pavlovskiy, ``Reducing over-smoothness in speech synthesis using generative adversarial networks,'' in \emph{2019 International Multi-Conference on Engineering, Computer and Information Sciences (SIBIRCON)}.\hskip 1em plus 0.5em minus 0.4em\relax IEEE, 2019, pp. 0972--0974.

\bibitem{kong2020hifi}
J.~Kong, J.~Kim, and J.~Bae, ``Hifi-gan: Generative adversarial networks for efficient and high fidelity speech synthesis,'' \emph{Advances in Neural Information Processing Systems}, vol.~33, pp. 17\,022--17\,033, 2020.

\bibitem{kavalerov2021multi}
I.~Kavalerov, W.~Czaja, and R.~Chellappa, ``A multi-class hinge loss for conditional gans,'' in \emph{Proceedings of the IEEE/CVF winter conference on applications of computer vision}, 2021, pp. 1290--1299.

\bibitem{kang2023libriheavy}
W.~Kang, X.~Yang, Z.~Yao, F.~Kuang, Y.~Yang, L.~Guo, L.~Lin, and D.~Povey, ``Libriheavy: a 50,000 hours asr corpus with punctuation casing and context,'' \emph{arXiv preprint arXiv:2309.08105}, 2023.

\bibitem{koizumi2023libritts}
Y.~Koizumi, H.~Zen, S.~Karita, Y.~Ding, K.~Yatabe, N.~Morioka, M.~Bacchiani, Y.~Zhang, W.~Han, and A.~Bapna, ``Libritts-r: A restored multi-speaker text-to-speech corpus,'' \emph{arXiv preprint arXiv:2305.18802}, 2023.

\bibitem{loshchilov2017decoupled}
I.~Loshchilov and F.~Hutter, ``Decoupled weight decay regularization,'' \emph{arXiv preprint arXiv:1711.05101}, 2017.

\bibitem{vaswani2017attention}
A.~Vaswani, N.~Shazeer, N.~Parmar, J.~Uszkoreit, L.~Jones, A.~N. Gomez, {\L}.~Kaiser, and I.~Polosukhin, ``Attention is all you need,'' \emph{Advances in neural information processing systems}, vol.~30, 2017.

\bibitem{mittag2021nisqa}
G.~Mittag, B.~Naderi, A.~Chehadi, and S.~M{\"o}ller, ``Nisqa: A deep cnn-self-attention model for multidimensional speech quality prediction with crowdsourced datasets,'' \emph{arXiv preprint arXiv:2104.09494}, 2021.

\bibitem{ravanelli2021speechbrain}
M.~Ravanelli, T.~Parcollet, P.~Plantinga, A.~Rouhe, S.~Cornell, L.~Lugosch, C.~Subakan, N.~Dawalatabad, A.~Heba, J.~Zhong \emph{et~al.}, ``Speechbrain: A general-purpose speech toolkit,'' \emph{arXiv preprint arXiv:2106.04624}, 2021.

\end{thebibliography}

\end{document}